\documentclass[aps,showpacs,preprint]{revtex4}
\usepackage[dvips]{graphics,graphicx}

\def\/{\over}
\def\<{\left\langle}
\def\>{\right\rangle}
\def\({\left(}
\def\){\right)}
\def\[{\left[}
\def\]{\right]}
\def\d{{\rm d}}

\def\e{{\rm e}}

\def\tr{\hbox{tr}}

\def\det{\hbox{det}\,}
\def\mod{\hbox{mod}\,}

\begin{document}
\title{Time-domain scars: resolving the spectral form factor in phase
space}
\author{Thomas Dittrich$^{1,2}$, Leonardo A.~Pach\'on$^{1,2}$}
\affiliation{$^{1}$Departamento de F\'{\i}sica, Universidad Nacional
de Colombia, Bogot\'a D.C., Colombia.\\
$^{2}$CeiBA -- Complejidad, Bogot\'a D.C., Colombia.}
\date{\today}
\begin{abstract}
We study the relationship of the spectral form factor with quantum as
well as classical probabilities to return. Defining a quantum return
probability in phase space as a trace over the propagator of the
Wigner function allows us to identify and resolve manifolds in phase
space that contribute to the form factor. They can be associated to
classical invariant manifolds such as periodic orbits, but also to
non-classical structures like sets of midpoints between periodic
points. By contrast to scars in wave functions, these features are not
subject to the uncertainty relation and therefore need not show any
smearing. They constitute important exceptions from a continuous
convergence in the classical limit of the Wigner towards the Liouville
propagator. We support our theory with numerical results for the
quantum cat map and the harmonically driven quartic oscillator.
\end{abstract}
\maketitle
\emph{Introduction} Evidence abounds that the spectrum of quantum
systems bears information on the corresponding classical dynamics, in
particular on manifolds invariant under time evolution. The Gutzwiller
trace formula \cite{gut71} and its numerous ramifications feature
specifically the set of isolated unstable periodic orbits of
classically chaotic systems. The discovery that energy eigenfunctions
are typically ``scarred'' along such orbits \cite{hel84} required to
modify the picture of ergodic eigenstates and allowed for the first
time to directly visualize the impact of classical invariant manifolds
on quantum mechanical distributions defined on configuration or phase
space \cite{TA&01}. 
The influence of classical invariant manifolds on time-domain features
has mainly been studied in the spectral form factor. It inherits its
relation to periodic orbits from the underlying spectral density via
the Gutzwiller trace formula. Being bilinear in the spectral density,
it involves pairs of orbits and their interfering contributions. A
host of research work has been dedicated to evaluating the double sum
over periodic orbits that ensues \cite{SR01}. Only recently, the full
sum could be tamed, thus providing an exact semiclassical account of
the form factor \cite{MH&04}. 

A step towards more global and immediate relationships to the
classical dynamics has been made in the context of the spectral
analysis of systems with dynamical localization \cite{DS91,dit96}, in
the form of a direct relation of the spectral form factor $K(\tau)$
\cite{DS91,dit96} with the classical probability to return $P_{\rm
ret}^{\rm cl}(t)$. For chaotic systems it reads
\begin{equation}\label{ffretprob}
K(\tau) \approx (2/\beta)\tau P_{\rm ret}^{\rm cl}(t_{\rm H}\tau),
\end{equation}
where $\beta = 1$ for systems invariant under time reversal and 2
otherwise. Being based on the diagonal approximation, the expression
is valid for times short compared to the Heisenberg time $t_{\rm
H}$. A similar relation holds for integrable systems, but without the
prefactor $\tau$. Equation (\ref{ffretprob}) calls for a deeper
understanding and analysis beyond its original application and
derivation from the Gutzwiller trace formula, to explore its potential
as an alternative semiclassical route to spectral analysis.

In this Letter, we study the relation of quantum and classical return
probabilities in phase space with the spectral form factor in the
light of recent progress in semiclassical approximations to the Wigner
propagator \cite{RO02,TDS06}. This approach has the special merit that
the interference of orbit pairs is already implicit in quantum return
probabilities. They can be expressed, like their classical analogues,
as traces (not traces squared!) over a corresponding propagator,
resulting in very direct quantum-classical relations on the same
footing.

Before tracing, the diagonal propagator of the Wigner function,
through its explicit dependence on phase-space coordinates, allows
to resolve the manifolds in phase space behind the contributions to
the form factor. Expressing it semiclassically in terms of orbit
pairs, it turns out that besides the classical invariant manifolds
also sets of midpoints between them contribute. Hence classical and
quantum return probabilities generally cannot coincide. This implies
severe restrictions to the convergence of the Wigner propagator
towards the classical (Liouville) propagator, at least for the
diagonal propagator near such midpoint manifolds. That these dominant
features of the diagonal Wigner propagator, classical as well as
non-classical ones, occur in a time-dependent distribution function
suggests calling them ``time-domain scars''. By contrast to scars in
eigenfunctions, they are not affected by the uncertainty relation and
therefore allow for an unlimited resolution of classical structures.

\emph{Classical and quantum return probabilities} In quantum
mechanics, a probability to return is generally defined like an
autocorrelation function: Introduce a return amplitude $a_{\rm ret}(t)
= \int{\rm d}^f q_0\langle {\bf q}(t)|{\bf q}_0\rangle$ with $|{\bf
q}(t)\rangle = \hat{U}(t) |{\bf q}_0\rangle$, $\hat{U}(t)$ the
time-evolution operator, and square, 
\begin{equation}\label{qretprobU}
P_{\rm ret}^{\rm qm}(t) = |a_{\rm ret}(t)|^2 = |\tr \hat{U}(t)|^2.
\end{equation}
By contrast, a classical return probability in phase space is
constructed as follows: Prepare a localized initial distribution
$\rho_{{\bf r}_0}({\bf r},0) = \delta_{\Delta}({\bf r}-{\bf r}_0)$,
$\delta_{\Delta}({\bf r})$ a strongly peaked function of width
$\Delta$ and ${\bf r} = ({\bf p},{\bf q})$ a vector in
$2f$-dimensional phase space. Propagate it over a time $t$ and overlap
it with the initial distribution. The resulting $p_{\rm
ret}^{\rm cl}({\bf r}_0,t) = \int{\rm d}^{2f}r\, \rho_{{\bf r}_0}({\bf
r},t) \rho_{{\bf r}_0}({\bf r},0)$ can be interpreted as a probability
density to return. Here, the time-evolved distribution is obtained
from the Liouville propagator $G^{\rm cl}({\bf r}'',t;{\bf r}',0)$ as
$\rho_{{\bf r}_0}({\bf r}'',t) = \int{\rm d}^{2f}r'\, G^{\rm cl}({\bf
r}'',t;{\bf r}',0) \rho_{{\bf r}_0}({\bf r}',0)$. Tracing over phase
space yields the return probability $P_{\rm ret}^{\rm cl}(t) =
\int{\rm d}^{2f}r_0\, p_{\rm ret}^{\rm cl}({\bf r}_0,t)$. Replacing
the initial distribution by $\delta({\bf r} - {\bf r}_0)$, we have
\begin{equation}\label{cretprob}
P_{\rm ret}^{\rm cl}(t) =
\int{\rm d}^{2f}r_0\, G^{\rm cl}({\bf r}_0,t;{\bf r}_0,0).
\end{equation}
To avoid divergences in particular at $t = 0$, the phase-space
integration has to be restricted to a finite range $\Delta E$ in
energy, if it is conserved, by introducing some normalized energy
distribution $\rho(E)$.


In quantum mechanics, the Wigner function allows for a similar
construction. Being related to the den\-si\-ty operator $\hat\rho(t)$
by an invertible transformation,
$W({\bf r},t)$ $=$ $\int\d^f q'\, \e^{-{\rm i}{\bf p\cdot q'}/\hbar}
\<{\bf q}+{\bf q}'/2\right|\hat\rho(t)\left|{\bf q}-{\bf q}'/2\>$,
its pro\-pa\-ga\-tor is defined as the kernel that evolves it over
finite time, $W({\bf r}'',t'') =$ $\int \d^{2f}r'\, G_{\rm W}({\bf
r}'',t'';{\bf r}',t') W({\bf r}',t')$. By analogy, we thus arrive at a
quantum-mechanical quasi-probability density to return in phase space
\cite{sar90}, $p_{\rm ret}^{\rm qm}({\bf r}_0,t) =$ $G_{\rm W}({\bf
r}_0,t;{\bf r}_0,0)$, and a return probability
\begin{equation}\label{qretprobG}
P_{\rm ret}^{\rm qm}(t) =
\int{\rm d}^{2f}r_0\, G_{\rm W}({\bf r}_0,t;{\bf r}_0,0).
\end{equation}
The integration across the energy shell produces a factor $D_{\cal{H}}
= \Delta E/\langle d\rangle$, the effective dimension of the Hilbert
space ${\cal{H}}$, $\langle d\rangle$ denoting the mean spectral
density.

Equations (\ref{qretprobG}) and (\ref{qretprobU}) are equivalent, as
becomes clear if we express the propagator of the Wigner func\-tion in
terms of the Weyl propagator, $U({\bf r},t) =$ $\int\d^f q'\,
\e^{-{\rm i}{\bf p\cdot q'}/\hbar}$ $\<{\bf q}+{\bf q}'/2\right|$
$\hat U(t)$ $\left|{\bf q}-{\bf q}'/2\>$,
\begin{equation}\label{wigweyl}
\!\!\!\!\!G_{\rm W}({\bf r}'',t;{\bf r}',0) = \int \d^{2f}r\,
\e^{{-{\rm i}\/\hbar}({\bf r}''-{\bf r}')\wedge{\bf r}}
U^*({\bf r}_-,t)U({\bf r}_+,t),
\end{equation}
with ${\bf r}_{\pm} \equiv ({\bf r}'+{\bf r}'' \pm {\bf r})/2$.
Substituting in Eq.~(\ref{qretprobG}) and transforming to ${\bf
r}'_{\pm} = {\bf r}_0 \pm {\bf r}/2$, the two integrals factorize,
$P_{\rm ret}^{\rm qm}(t) = \int{\rm d}^{2f}r'_- U^*({\bf r}'_-,t)
\int{\rm d}^{2f}r'_+ U({\bf r}'_+,t)$ $=|\tr \hat U(t)|^2$.

\emph{Form factor and diagonal propagator} Also the form factor is
related to the trace-squared of the time-evolution operator,
$K(t/t_{\rm H}) = D_{\cal{H}}^{-1}$ $|\tr \hat U(t)|^2$ for $t\gtrsim
t_{\rm H}/D_{\cal{H}}$, where $t_{\rm H} = h\langle d\rangle$. The
factor $D_{\cal{H}}^{-1}$ normalizes $\lim_{\tau\to\infty}K(\tau) =
1$. By comparison with Eqs.~(\ref{qretprobU}) and (\ref{qretprobG}),
\begin{equation}\label{propformfac}
P_{\rm ret}^{\rm qm}(t) =
\int{\rm d}^{2f}r\, G_{\rm W}({\bf r},t;{\bf r},0) = 
D_{\cal{H}} K(t/t_{\rm H}).
\end{equation}
This remarkable relation expresses the form factor as the trace over a
quantity with a close classical analogue, not as a squared trace. It
is an exact identity and does not involve any semiclassical
approximation.

Contrast Eq.~(\ref{propformfac}) with (\ref{ffretprob}). Both relate
$K(\tau)$ with a return probability, but there is a clear discrepancy,
manifest in the factor $\tau$ that appears only in (\ref{ffretprob}).
This may not be surprising given that the two relations refer to
return probabilities on the quantum and the classical level,
respectively. However, if we take into account also
Eqs.~(\ref{cretprob}) and (\ref{qretprobG}), we face a
dilemma: There is ample evidence \cite{mcl83,RO02,TDS06} that the
Wigner propagator generally converges in the classical limit to the
Liouville propagator,
\begin{equation}\label{qpropcprop}
\lim_{\hbar\to 0} G_{\rm W}({\bf r}'',t;{\bf r}',0) =
G^{\rm cl}({\bf r}'',t;{\bf r}',0).
\end{equation}
For up to quadratic Hamiltonians, is even identical to it. Were
Eq.~(\ref{qpropcprop}) correct also for ${\bf r}' = {\bf r}''$---and
on the diagonal the Wigner propagator should behave \emph{more
classically} than elsewhere---then $\lim_{\hbar\to 0} P_{\rm ret}^{\rm
qm}(t) = P_{\rm ret}^{\rm cl}(t)$ should hold as well!

The derivation of Eq.~(\ref{ffretprob}) \cite{DS91,dit96} suggests
that the factor $\tau$ arises as a degeneracy factor due to the
coherent superposition of contributions from different points along a
given periodic orbit, each of which can be interpreted as a periodic
point of its own, $\tau$ measuring the magnitude of this set in phase
space. We therefore suspect that Eq.~(\ref{qpropcprop}) might fail in
the presence of constructive quantum interference. This can be
substantiated taking into account semiclassical approximations for
$G_{\rm W}({\bf r}'',t;{\bf r}',0)$ based on \emph{pairs} of classical
trajectories \cite{RO02,TDS06} ${\bf r}_-^{\rm cl}(t)$, ${\bf
r}_+^{\rm cl}(t)$, chosen such that for their respective initial
points ${\bf r}'_\pm$, ${\bf r}' = ({\bf r}'_- + {\bf r}'_+)/2$, and
likewise for ${\bf r}''_\pm$. Specifically for the diagonal
propagator, this requires that both ${\bf r}_-^{\rm cl}(t)$ and ${\bf
r}_+^{\rm cl}(t)$ be periodic orbits. The set of midpoints $\bar{\bf
r}(t) = ({\bf r}_-^{\rm cl}(t) + {\bf r}_+^{\rm cl}(t))/2$ then forms
a closed curve in phase space as well and contributes to the diagonal
propagator hence the form factor, but \emph{need not consist of
periodic points proper}.

It is tempting to interpret also the prefactor $2/\beta$ in
Eq.~(\ref{ffretprob}) as a degeneracy factor and to look for 
phase-space manifolds that in time-reversal invariant systems
contribute the extra weight to $P_{\rm ret}^{\rm qm}(t)$: They
can be found in sets of midpoints between symmetry-related pairs of
periodic orbits, located in the symmetry (hyper)plane ${\bf p} =
0$. Similarly, other non-diagonal contributions to the form factor
\cite{SR01,MH&04} can be associated to non-classical enhancements of
the diagonal Wigner propagator.

\begin{figure}[floatfix]
\centerline{\includegraphics[width=5cm,angle=0]{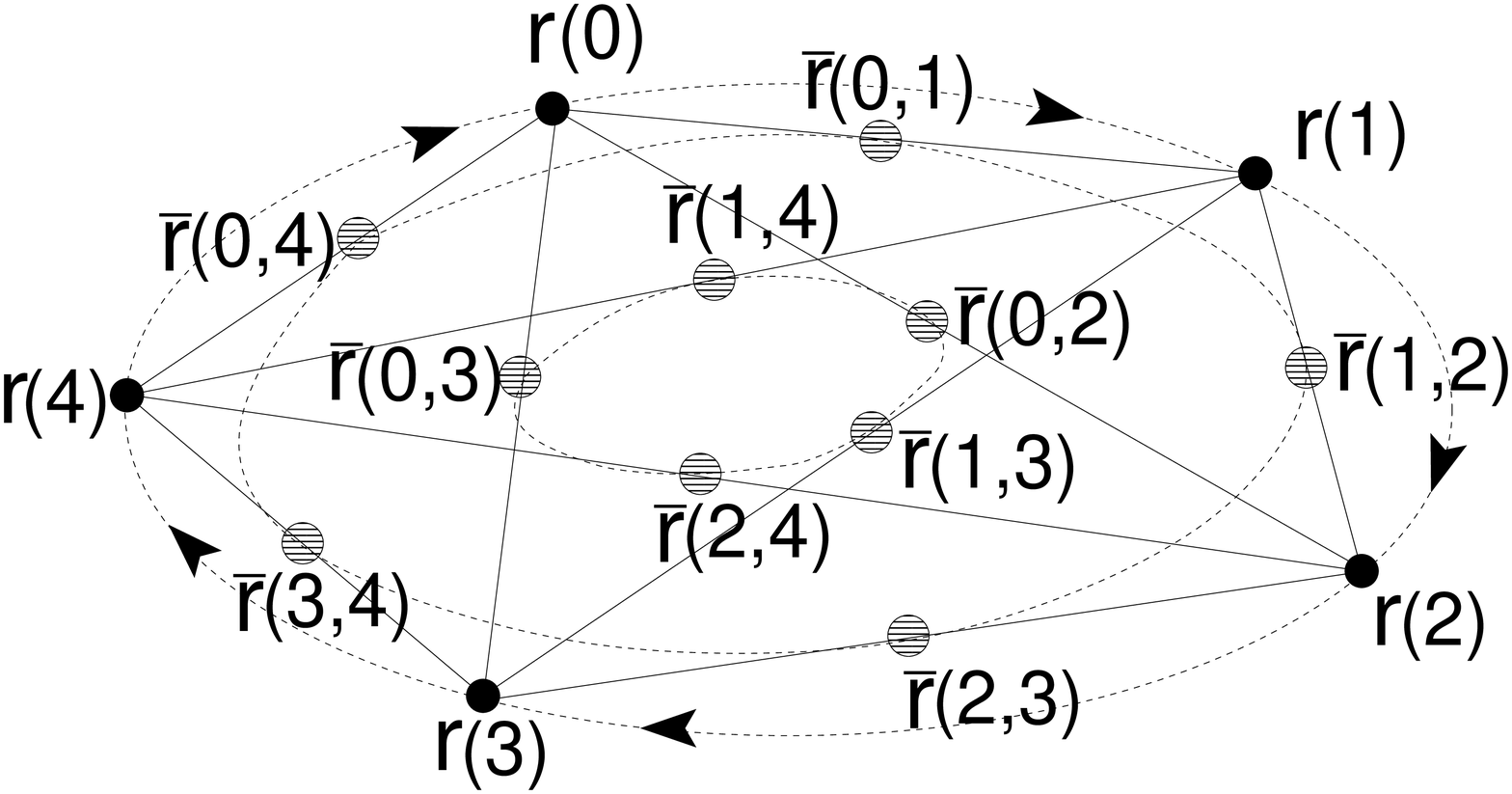}}
\vspace*{-0.3cm}
\caption{\label{f_midpoints} 
Schematic drawing of a set of periodic points with period 5 of a
symplectic map with their midpoints.
}
\end{figure}

\emph{Examples} In order to render our argument more quantitative, we
first discuss the case of discrete time: Consider a set of periodic
points ${\bf r}_j(n + N_j) = {\bf r}_j(n)$, $n = 0,\ldots,N_j-1$, of a
symplectic map $\cal{M}$. In their vi\-ci\-ni\-ty, the semiclassical
Wigner propagator is given by $G_{{\rm W}j}({\bf r}'',N_j;{\bf r}',0)
= \delta({\bf r}'' - M_j{\bf r}')$, $M_j$ denoting ${\cal{M}}^{N_j}$
linearized near ${\bf r}'$, ${\bf r}''$. Define midpoints $\bar{\bf
r}_j(m,n) = ({\bf r}_j(m) + {\bf r}_j(n))/2$ (cf.\
Fig.~\ref{f_midpoints}). By construction, $\bar{\bf r}_j(m + N_j,n) =
\bar{\bf r}_j(m,n)$, but generally ${\cal{M}}^{N_j}\bar{\bf r}_j(m,n)
\neq \bar{\bf r}_j(m,n)$. For ${\bf r}' \approx {\bf r}'' \approx
\bar{\bf r}_j(m,n)$, the Wigner propagator carries an additional
oscillatory factor,
\begin{eqnarray}\label{wignerpp}
&&G_{{\rm W}j}({\bf r}'',N_j;{\bf r}',0) =
2\delta({\bf r}'' - M_j{\bf r}')\times \nonumber\\
&&\cos\big(({\bf r}_j(n)-{\bf r}_j(m))\wedge
({\bf r}''-{\bf r}')/\hbar\big).
\end{eqnarray}
From here, tracing reduces to equating ${\bf r}'$ with ${\bf r}''$ and
summing points. There are $N_j$ periodic points on the orbit and
$N_j(N_j-1)$ midpoints ($\bar{\bf r}_j(m,n)$ and $\bar{\bf r}_j(n,m)$
count separately), resulting in a total return probability
\begin{equation}\label{qretproppp}
P_{{\rm ret}\,j}^{\rm qm}(N_j) = 
N_j^2/|\det(M_j-I)| = N_j P_{{\rm ret}\,j}^{\rm cl}(N_j).
\end{equation}
The midpoints' contribution thus is responsible for the extra factor
$\tau$, i.e. here, $N_j$ and explains the discrepancy between
classical and quantum return probabilities.

\begin{figure}[floatfix]
\centerline{\includegraphics[width=3.7cm,angle=270]{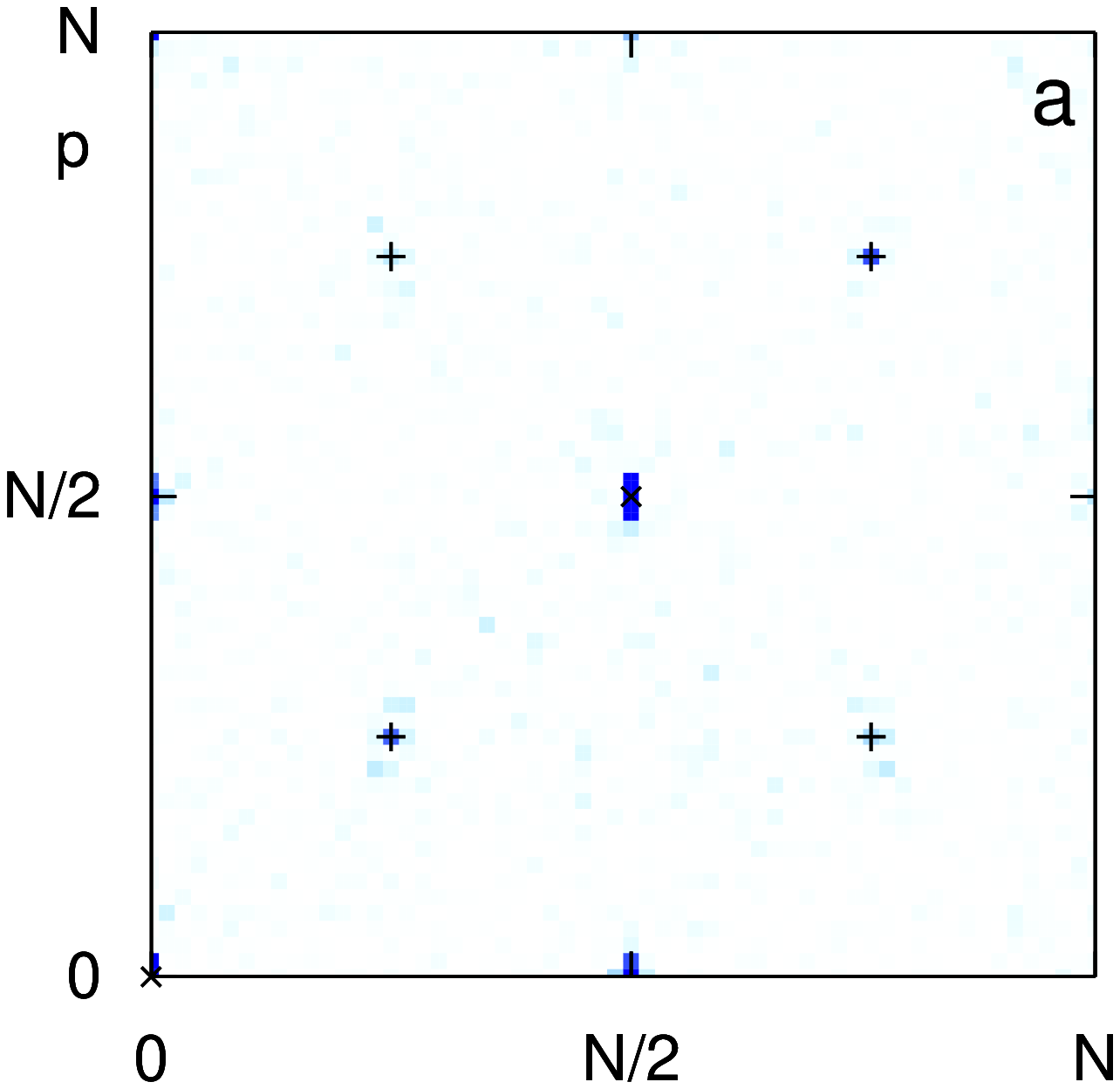}
            \hspace*{-0.4cm}
            \includegraphics[width=3.7cm,angle=270]{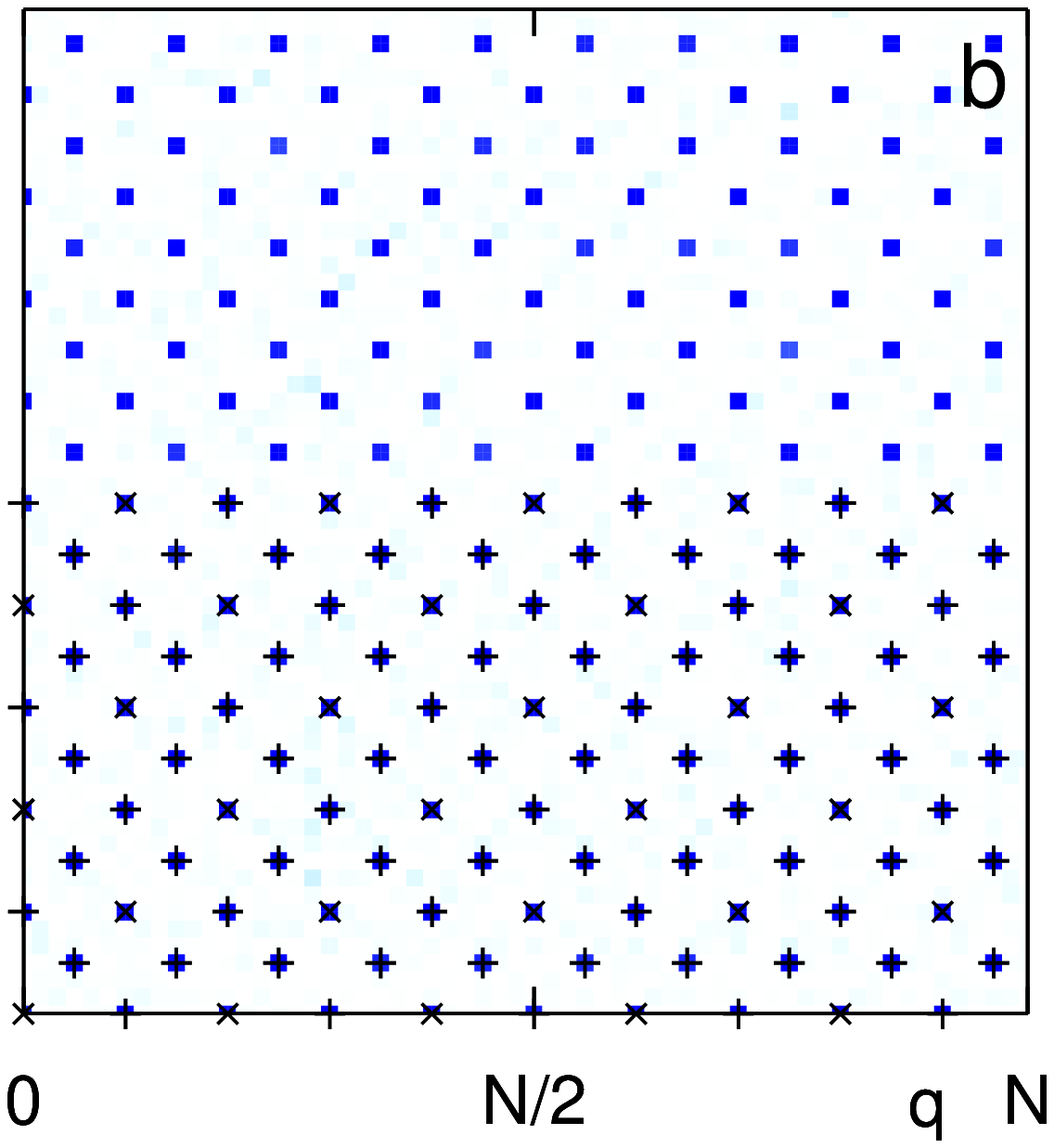}}
\vspace*{-0.4cm}
\caption{\label{f_arnold} 
Diagonal Wigner propagator $G_{\rm W}({\bf r},n;{\bf r},0)$ for the
quantized Arnol'd cat map at $n = 1$ (a) and $n = 3$ (b). Symbols
$\times$, $+$ mark periodic points of the corresponding classical map
and their midpoints, respectively (for better visibility of the data,
symbols have been suppressed in the upper half of panel (b)). The
Hilbert-space dimension is $D_{\cal{H}} = 60$. Color code ranges from
red (negative) to blue (positive).
}
\end{figure}

As an example, consider the Arnol'd cat map. It is defined on a torus,
${\bf r}'' = T {\bf r}' (\mod 1)$, ${\bf r} \in [0,1)^2$, $T$ a
$2\times 2$ matrix with integer coefficients. We choose the simplest
combination that allows for quantization \cite{HB80}, $T =
(2,1;3,2)$. The topology of the underlying classical space implies
that both position and momentum be quantized, leading to a finite
Hilbert-space dimension $D_{\cal{H}}$. The definition of the Wigner
function can be adapted to this discrete periodic Hilbert space to
avoid redundancies \cite{AB95,AD05}. In Fig.~\ref{f_arnold}, we show
the diagonal Wigner propagator after 1 and 3 iterations of the quantum
map. The peaks of the diagonal propagator coincide perfectly with the
periodic points of the classical map. Moreover, they appear with
almost \emph{single-pixel precision}. While the uncertainty relation
requires a minimum area of $D_{\cal{H}}$ pixels, this is perfectly
admissible for the propagator. To check Eq.~(\ref{qretproppp}), we
compared the trace of the diagonal propagator to analytical results
for $\sum_j N_j^2/|\det(M_j-I)|$ (2.0 and 50.0, resp.), and found
coincidence up to $6$ digits.

\begin{figure}[floatfix]
\centerline{\includegraphics[width=6.5cm,angle=0]{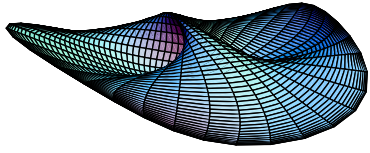}}
\vspace*{-0.4cm}
\caption{\label{f_caustic} 
Surface formed by midpoints of a fictitious periodic orbit that is not
circularly symmetric nor confined to a plane in phase space. It
exhibits self-intersections but retains the topology of a closed
two-dimensional ribbon, see text.
}
\end{figure}

\begin{figure}[floatfix]
\centerline{\includegraphics[width=5.3cm,angle=270]{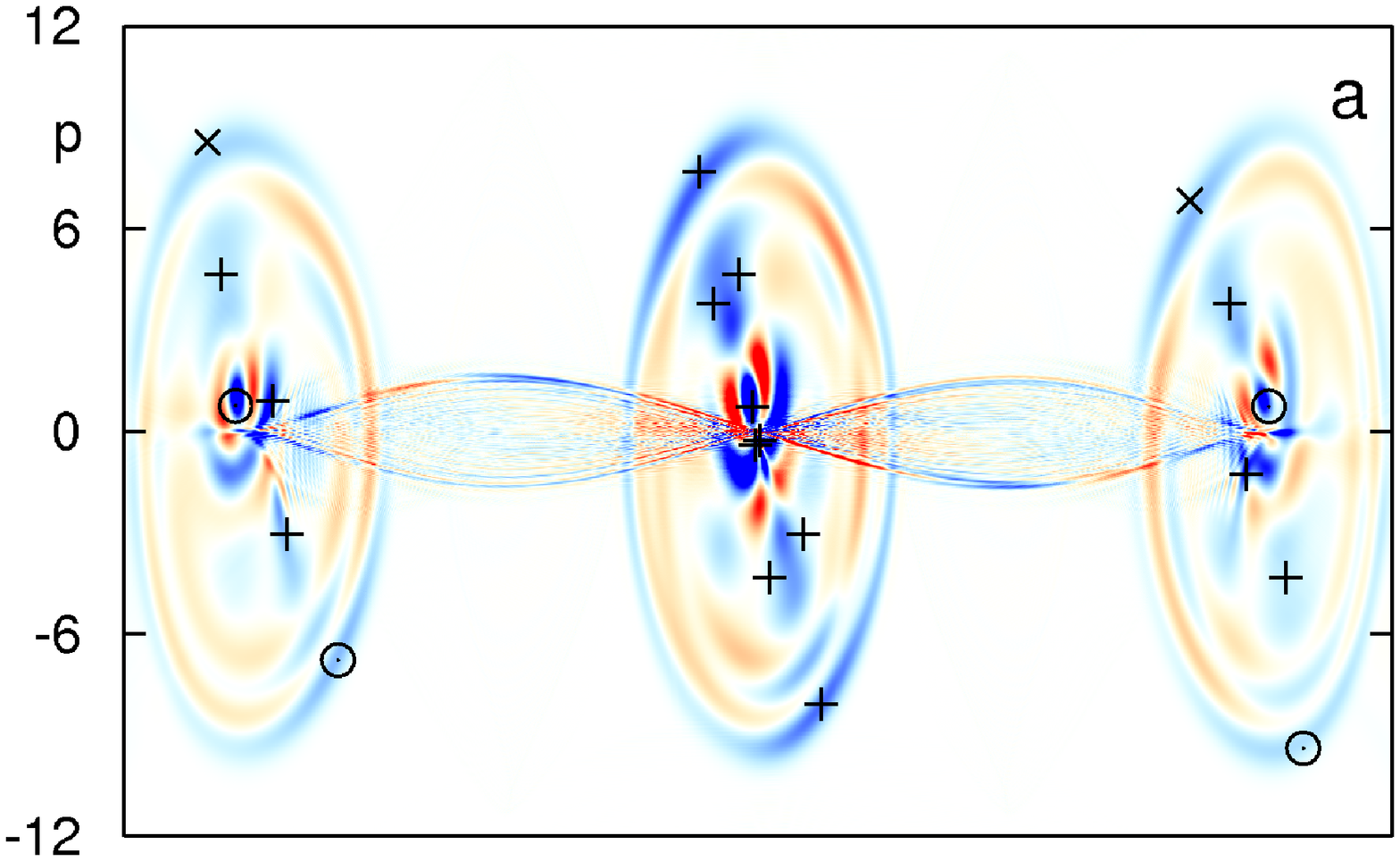}}
\vspace*{-0.4cm}
\hspace*{-0.1cm}
\centerline{\includegraphics[width=5.3cm,angle=270]{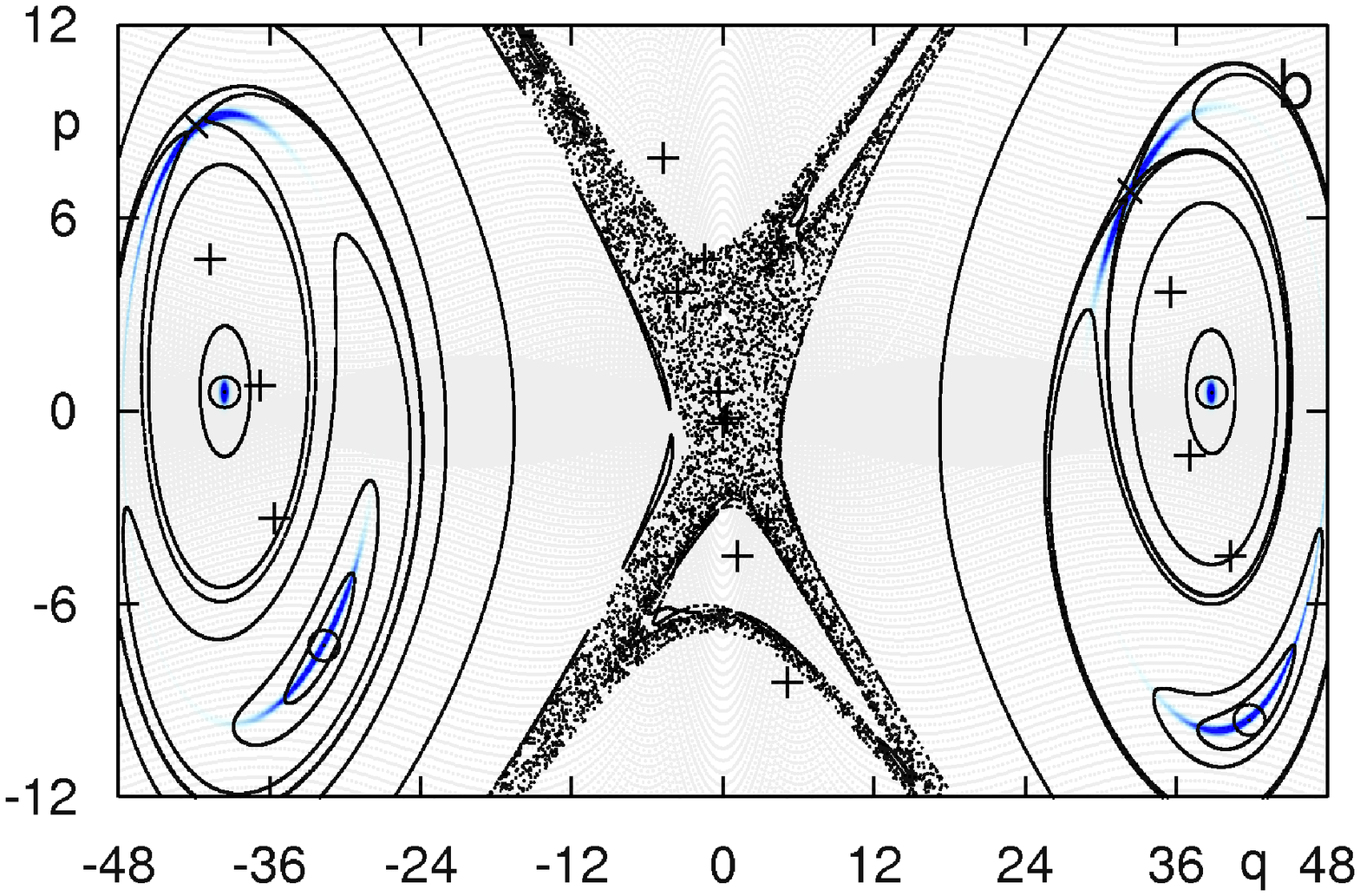}}
\vspace*{-0.cm}
\caption{\label{f_hdqo} 
Diagonal Wigner (a) and Liouville (b) propagators $G({\bf r},t;{\bf
r},0)$ for the harmonically driven quartic oscillator at $t = T \equiv
2\pi/\omega$, with $\omega_0 = 1.0$, $\omega = 0.95$, $\phi = \pi/3$,
$S = 0.07$, and $E_{\rm b} = 192.0$ (color code as in
Fig.~\protect\ref{f_arnold}). For better orientation, we superimpose a
stroboscopic surface of section of the same system (panel
(b), black). The figure-$\infty$ structure is the Wigner caustic of a
period-$T$ torus outside the frame shown (grey). Symbols $\odot$,
$\times$ mark elliptic and hyperbolic periodic points of the classical
system, resp., and $+$ their midpoints.
}
\end{figure}

Going to systems in continuous time, a periodic orbit ${\bf r}_j(s) =
{\bf r}_j(s + T_j)$ gives rise to midpoints $\bar{\bf r}_j(s',s'') =
({\bf r}_j(s') + {\bf r}_j(s''))/2$. This replaces
Eq.~(\ref{wignerpp}) with
\begin{eqnarray}\label{wignerpo}
&&G_{{\rm W}j}({\bf r}'',t;{\bf r}',0) =
2\delta({\bf r}'' - M_j{\bf r}')\times \nonumber\\
&&\cos\big(({\bf r}_j(s'')-{\bf r}_j(s'))\wedge
({\bf r}''-{\bf r}')/\hbar\big) \delta(t-T_j).
\end{eqnarray}
The midpoints now merge into a continuous two-di\-men\-sio\-nal
surface ${\cal{S}}_j$ parameterized by $(s',s'')$, $0 \leq s',s'' <
T_j^{\rm p}$, the length of the orbit. Topologically it forms a closed
ribbon. As a consequence, the diagonal propagator consists of a
$\delta$-function only in the subspace orthogonal to ${\cal{S}}_j$,
$G_{{\rm W}j}({\bf r},t;{\bf r},0) = \delta({\bf r}_{\perp})
\delta(t-T_j)/|\det(M_{j\perp}-I)|$, where $M_{j\perp}$ is the
stability matrix restricted to the $(2f-2)$-dimensional subspace ${\bf
r}_{\perp}$. Upon tracing, the integration over ${\cal{S}}_j$ yields a
factor ${T_j^{\rm p}}^2$, its effective area,
\begin{equation}\label{qretproppo}
P_{{\rm ret}\,j}^{\rm qm}(t) = 
\Delta E\,{T_j^{\rm p}}^2\,\delta(t-T_j)/2\pi\hbar |\det
(M_{j\perp}-I)|.
\end{equation}
In Cartesian phase-space coordinates ${\bf r}$, ${\cal{S}}_j$
may have a nontrivial geometry. In general, it will exhibit a Wigner
caustic \cite{ber76}, an overlap of three leaves near the center of
the orbit, owing to the fact that a given point in this region may be
the midpoint of more than one pair of periodic points on the orbit. 
The phenomenon can well be observed in Fig.~\ref{f_hdqo}. If the
periodic orbit is not confined to a plane, this geometric degeneracy
will be lifted, resulting in folds and self-intersections, illustrated
in Fig.~\ref{f_caustic} for a fictitious periodic orbit.

A pertinent example is the harmonically driven quar\-tic oscillator
$H(p,q,t) = p^2/2m - m\omega_0^2q^2/4 + m^2\omega_0^4q^4/64E_{\rm b} +
Sq\cos(\omega t + \phi)$ \cite{GD&91}, with generally mixed phase
space. %
In the diagonal propagator at $t = T \equiv 2\pi/\omega$
(Fig.~\ref{f_hdqo}) we identify a number of isolated peaks at periodic
points of the classical dynamics, elliptic as well as hyperbolic, and
their midpoints, and an enhancement over a well-defined region, to be
interpreted as the Wigner caustic of a period-$T$ torus outside the
frame shown, as confirms the coincidence with the corresponding
classical feature in Fig.~\ref{f_hdqo}b.

\emph{Refinements and perspectives} An alternative access to the
Wigner propagator near periodic orbits is Berry's scar function, a
semiclassical approximation to the Weyl propagator in the energy
domain \cite{ber89}. It responds to the special situation close to a
periodic orbit $j$ by using local curvilinear coordinates: energy,
time, and remaining phase-space directions ${\bf r}_{j\perp}$
perpendicular to the orbit. Transformed to the time domain and
substituted for the Weyl propagator in Eq.~(\ref{wigweyl}), it leads
to a semiclassical approximation for the diagonal Wigner propagator,
\begin{equation}\label{berrypo}
G_{{\rm W}j}({\bf r},t;{\bf r},0) =
\frac{T_j^{\rm p}/2\pi\hbar}{|\det(M_{j\perp}-I)|}
\delta({\bf r}_{j\perp})\delta(t-T_j).
\end{equation}
The primitive period $T_j^{\rm p}$ and the determinantal prefactor
measure the length and the effective cross section, resp., of the
``phase-space tube'' around the orbit that contributes to the diagonal
propagator. By contrast to Eq.~(\ref{wignerpo}), the degeneracy factor
$T_j^{\rm p}$ appears here already before tracing: The use of
local coordinates condenses the contributions of periodic points as
well as midpoints onto the orbit. Equation (\ref{berrypo}) does not
apply outside the orbit $j$ and therefore does not allow for
indiscriminate tracing over all of phase space.

The midpoint contribution to $G_{\rm W}({\bf r},t;{\bf r},0)$ giving
rise to marked non-classical features is a manifestation of quantum
coherence. It measures the quantum return probability for
Schr\"odinger-cat states distributed over different points of the same
periodic orbit. In the presence of incoherent processes, it decays on
the dephasing timescale. The Wigner propagator, operating on the
projective Hilbert space, readily permits including this effect
\cite{zur01} and thus to identify exclusively the classical invariant
manifolds, unaffected by the uncertainty relation, as peaks of a
purely quantum-mechanical distribution. Phase-space features
associated to non-diagonal contributions to the form factor will be
even more elusive and geometrically more involved, but are in
principle accessible to numerical study.

We have provided analytical and numerical evidence that
Eq.~(\ref{ffretprob}) can be interpreted as a global relation between
quantum and classical return probabilities which can be broken down
into contributions of invariant phase-space manifolds. They enter with
weight factors that measure the size of the set contributing
coherently, and lead to important exceptions to
Eq.~(\ref{qpropcprop}). Analytical evidence based on presently
available semiclassical approximations \cite{TDS06} indicates they
are restricted to the diagonal ${\bf r}' = {\bf r}''$ (where they are
least expected) and hence of measure zero. They are qualitatively
different for integrable systems: In action-angle variables, the size
of the degenerate sets is independent of time \cite{dit96} and
therefore does not contribute an extra factor $t$. This in turn
reflects the different dimensions and topologies of periodic tori vs.\
isolated unstable periodic orbits, indicating how to generalize this
to more involved cases like systems with mixed phase space. Merging
the different contributions on the classical side into more global
quantities like the Frobenius-Perron modes \cite{AA&96} remains as a
challenge for future research.

Fruitful discussions with D.~Braun, F.~Haake, H.~J.~Korsch, A.\ M.\
Ozorio de Al\-mei\-da, T.~H.~Seligman, M. Sieber, U.~Smilansky,
R.~Vallejos, and financial support by Colciencias, U.\ Nal.\ de
Colombia, and Volkswagen\-Stiftung are acknowledged with pleasure. We
enjoyed the hospitality extended to us by CBPF (Rio de Janeiro), CIC
(Cuernavaca), MPIPKS (Dresden), and U.\ of Technology Kaiserslautern.

\bibliographystyle{unsrt}
\bibliography{tdslet}

\begin{thebibliography}{10}

\bibitem{gut71}
M.~C. Gutzwiller.
\newblock {\em J.~Math.~Phys.}, 12:343, 1971.

\bibitem{hel84}
E.~J. Heller.
\newblock {\em Phys.~Rev.~Lett.}, 53:1515, 1984.

\bibitem{TA&01}
F.~Toscano {\em et al.}
\newblock {\em Phys.~Rev.~Lett.}, 86:59, 2001.

\bibitem{SR01}
M.~Sieber {\em et al.}
\newblock {\em Physica Scripta}, T90:128, 2001.

\bibitem{MH&04}
S.~M\"uller {\em et al.}
\newblock {\em Phys.~Rev.~Lett.}, 93:014103, 2004.

\bibitem{DS91}
T.~Dittrich and U.~Smilansky.
\newblock {\em Nonlinearity}, 4:85, 1991.

\bibitem{dit96}
T.~Dittrich.
\newblock {\em Phys.~Rep.}, 271:267, 1996.

\bibitem{RO02}
P.~P. de~M.~Rios~{\em et al.}
\newblock {\em J.~Phys.~A}, 35:2609, 2002.

\bibitem{TDS06}
T.~Dittrich {\em et al.}
\newblock {\em Phys.~Rev.~Lett.}, 96:070403, 2006.

\bibitem{sar90}
M.~Saraceno.
\newblock {\em Ann.~Phys.~(NY)}, 199:37, 1990.

\bibitem{mcl83}
F.~McLafferty.
\newblock {\em J.~Chem.~Phys.}, 78:3253, 1983.

\bibitem{HB80}
J.~H. Hannay and M.~V. Berry.
\newblock {\em Physica}, 1D:267, 1980.

\bibitem{AB95}
O.~Agam and N.~Brenner.
\newblock {\em J.~Phys.~A}, 28:1345, 1995.

\bibitem{AD05}
A.~Arg\"uelles and T.~Dittrich.
\newblock {\em Physica A}, 356:72, 2005.

\bibitem{ber76}
M.~V. Berry.
\newblock {\em Phil.~Trans.~Roy.~Soc.~A}, 387:237, 1976.

\bibitem{GD&91}
F.~Grossmann {\em et al.}
\newblock {\em Phys.~Rev.~Lett.}, 67:516, 1991.

\bibitem{ber89}
M.~V. Berry.
\newblock {\em Proc.~Roy.~Soc.~Lond.~A}, 423:219, 1989.

\bibitem{zur01}
W.~H. Zurek.
\newblock {\em Nature}, 412:712, 2001.

\bibitem{AA&96}
A.~V.~Andreev {\em et al.}
\newblock {\em Phys.~Rev.~Lett.}, 76:3947, 1996.

\end{thebibliography}

\end{document}